\documentstyle[buckow]{article}
\begin {document}
\def\titleline{
Open Branes and Little Strings
}
\def\authors{
Troels Harmark
}
\def\addresses{
Niels Bohr Institute, Blegdamsvej 17, 2100 Copenhagen, Denmark.
}
\def\abstracttext{
This is a short review of the newly discovered OD$p$-theories
that are non-gravitational six-dimensional theories
defined as the decoupling limit of NS5-branes in the
presence of a near-critical $(p+1)$-form RR fields.
We discuss the motivation for these new theories,
their definitions and properties, and their relation to
NCOS theory, OM theory and Little String Theory, 
focusing on the cases $p=1,2$.
}
\large
\makefront

\newcommand{\bls}{\bar{l}_s}
\newcommand{\bgs}{\bar{g}_s}
\newcommand{\los}{l_{\rm os}}
\newcommand{\lom}{l_{\rm om}}
\newcommand{\eps}{\varepsilon}
\newcommand{\spa}{\ \ , \ \ }
\newcommand{\gym}{g_{\rm YM}}
\newcommand{\tlst}{T_{\rm LST}}
\newcommand{\tncos}{T_{\rm NCOS}}
\newcommand{\eqref}[1]{(\ref{#1})}
\newcommand{\nf}{N_{\rm F1}}
\newcommand{\ndp}{N_{{\rm D}p}}
\newcommand{\nmt}{N_{\rm M2}}
\newcommand{\nmf}{N_{\rm M5}}
\newcommand{\leff}{l_{\rm eff}}

\newcommand{\href}[1]{}

\section{Introduction}

This talk is based mainly on the paper \cite{Harmark:2000ff}
and in addition on the 
papers \cite{Harmark:2000hw,Harmark:2000wv,Harmark:2000qm}.
The purpose of this talk is to review the key arguments and results
of the paper \cite{Harmark:2000ff}. The talk was given at the RTN
workshop ``The quantum structure of spacetime and the geometric nature
of fundamental interactions''
in Berlin, October 4-10, 2000.

Recently, new consistent supersymmetric theories have been discovered
as limits of branes in the presence of near-critical electric fields.
In \cite{Seiberg:2000ms,Gopakumar:2000na} 
the $p+1$ dimensional non-commutative open string (${\rm NCOS}_{p+1}$) 
theory was 
discovered as a decoupling limit of D$p$-branes with a near-critical
electrical NSNS two-form potential.
In \cite{Gopakumar:2000ep,Bergshoeff:2000ai} 
OM theory was discovered as a decoupling limit of the M5-brane
with a near-critical electric three-form potential, 
and finally in \cite{Harmark:2000ff,Gopakumar:2000ep} 
the OD$p$-theories\footnote{The OD1 and OD2-theories
are called $(1,1)$ and $(2,0)$ OBLSTs in \cite{Harmark:2000ff},
as we discuss below.} 
were discovered as decoupling limits of the NS5-brane with near-critical
electrical RR $(p+1)$-form potentials.

The interesting property that all these non-gravitational theories 
share is that 
they all have phases were the effective dynamics are governed by
massive dynamical extended objects. 
Prior to the discovery of these new theories, the
Little String Theory (LST) \cite{Berkooz:1997cq,Seiberg:1997zk} 
(see also \cite{Dijkgraaf:1997cv,Dijkgraaf:1997hk}) 
was the only non-gravitational theory with this property.
The hope is that these theories, being non-gravitational, eventually
can be understod indendently of the critical string theories
(this is already true for NCOS theories \cite{Gopakumar:2000na}).
This might then eventually improve the understanding of string theory
itself.

In this short review we briefly discuss NCOS theory and OM theory,
and then move to the definition and properties of $(1,1)$ OBLST/OD1-theory,
including the thermodynamics and the phase structure.
We subsequently consider the $(2,0)$ OBLST/OD2-theory and finally
we briefly look at other OD$p$-theories.

\section{Short review of NCOS and OM theory}

\subsection{NCOS theory}
\label{NCOSsec}

NCOS theory was discovered in \cite{Seiberg:2000ms,Gopakumar:2000na}.
${\rm NCOS}_{p+1}$ is obtained from the F1-D$p$ bound state, i.e. a bound 
state of F-strings dissolved in a D$p$-brane.
We take the F-strings to be charged in the $x^1$ direction and
the D$p$-brane to be charged in the $x^1$ to $x^p$ directions.

The ${\rm NCOS}_{p+1}$ decoupling limit is%
\footnote{We use the ${\rm NCOS}_{p+1}$ limit used in \cite{Gopakumar:2000na}
rather than the one used in \cite{Gopakumar:2000ep}.}
\begin{equation}
\label{F1Dplimit1}
l_s \rightarrow 0
\spa
\eps = \frac{l_s^2}{\los^2}
\spa
g_s = \tilde{g} \eps^{-1}
\end{equation}
\begin{equation}
\label{F1Dplimit2}
g_{\mu \nu} = \eps^{-1} \eta_{\mu \nu} 
\spa
\mu,\nu = 0,1
\spa
g_{ij} = \eps \delta_{ij} 
\spa
i,j = 2,...,9
\end{equation}
where $l_s$ is the closed string length, $g_s$ is the closed string coupling
and $g_{\mu\nu}$ is the closed string metric.

The limit \eqref{F1Dplimit1}-\eqref{F1Dplimit2} gives the
NSNS two-form field $B_{\mu \nu}$ and RR $(p-1)$-form field 
$A_{\mu_1 \cdots \mu_{p-1}}$
\begin{equation}
B_{01} = \frac{1}{\eps} \frac{1}{2\pi \bls^2} - \frac{1}{4\pi \los^2}
\spa 
A_{2\cdots p} = \frac{1}{(2\pi)^{p-2} \tilde{g} \los^{p-1} }
\end{equation}
This can be obtained using the Seiberg-Witten relations between closed
and open string moduli \cite{Seiberg:1999vs} 
together with the duality relation between $B_{01}$ and $A_{2\cdots p}$.

If we have $\nf$ F-strings and $\ndp$ D$p$-branes in the bound state
we have the RR $(p-1)$-form field from magnetic flux quantization
\( A_{2\cdots p} = \frac{2\pi}{\ndp} \frac{\nf}{V_{p-1}} \).
where $V_{p-1}$ is the volume of the space with coordinates 
$x^2$ to $x^p$. If $V_{p-1}$ is infinite, 
we still have $\frac{\nf}{V_{p-1}}$
finite.
From this we get the NCOS coupling
\( \tilde{g} = \frac{V_{p-1}}{(2\pi \los)^{p-1}} \frac{\ndp}{\nf} \).

We can see in a simple way that open F-string excitations can be
light in the background field $B_{01}$, since the effective tension is
\( T_{\rm eff} = \frac{1}{\eps} \frac{1}{2\pi l_s^2} - B_{01}
= \frac{1}{4\pi \los^2} \).
The ${\rm NCOS}_{p+1}$ is a non-gravitational theory
of open strings with only one orientation, and no closed strings
when the $x^1$ directions is non-compact.
From the Seiberg-Witten relations between closed and open string
moduli \cite{Seiberg:1999vs}
it can be seen that we have space-time non-commutativity
\( [x^0,x^1] = i 2\pi \los^2 \).
The theory reduces to a $p+1$ dimensional super Yang-Mills theory
(${\rm SYM}_{p+1}$) at low energies
with Yang-Mills coupling \( \gym^2 = (2\pi)^{p-2} \tilde{g} \los^{p-3} \).
The thermodynamics of NCOS theories have been considered in 
\cite{Harmark:2000wv,Klebanov:2000pp,Sahakian:2000ay,Gubser:2000mf}.

\subsection{OM theory}

OM theory was discovered in \cite{Gopakumar:2000ep,Bergshoeff:2000ai}.
OM theory is obtained from the M2-M5 bound state which is 
a bound state of M2-branes dissolved in an M5-brane.
We take the M2-branes to be charged in the $x^1$ and $x^2$ directions and
the M5-brane to be charged in the $x^1$ to $x^5$ directions.
The OM theory decoupling limit is then
\begin{equation}
\label{OMlimit1}
l_p \rightarrow 0
\spa
f = \frac{l_p^6}{\lom^6}
\end{equation}
\begin{equation}
\label{OMlimit2}
g_{\mu \nu} = f^{-1} \eta_{\mu \nu} 
\spa
\mu,\nu = 0,1,2
\spa
g_{ij} = f \delta_{ij} 
\spa
i,j = 3,...,10
\end{equation}
Using magnetic flux quantization and the self-duality of the 
three-form field strength on the M5-brane we get the three form potential
\begin{equation}
C_{012} = \frac{1}{f} \frac{1}{(2\pi)^2 l_s^3} - \frac{1}{2(2\pi)^2 \lom^3}
\spa
C_{345} = \frac{1}{(2\pi)^2 \lom^3}
\end{equation}
along with $C_{345} = \frac{2\pi}{\nmf} \frac{\nmt}{V_3}$
and $\frac{\nmf}{\nmt} \frac{V_3}{(2\pi \lom)^3} = 1$
where $V_3$ is the volume of the space given by the 
$x^3$ to $x^5$ directions and where $\nmt$ and $\nmf$ are the
number of M2 and M5-branes, respectively.

Open M2-branes are light in the background of the $C_{012}$
field when suitably oriented, since the effective
membrane tension is
\( T_{\rm eff} = \frac{1}{f} \frac{1}{(2\pi)^2 l_p^3} - C_{012}
= \frac{1}{2(2\pi)^2 \lom^3} \).
OM theory is thus a non-gravitational theory of open membranes,
that for \( \lom \rightarrow 0 \) reduces to the $(2,0)$ SCFT.

When compactifying OM theory on an electric circle we get
${\rm NCOS}_{4+1}$ \cite{Gopakumar:2000ep,Bergshoeff:2000ai}.
An electric circle means a circle in one of the directions that the
M2-branes are charged. Since this procedure requires identification
of remaining part of the 11-dimensional metric with the 10-dimensional
metric we require that \( \eps = f \).
From this we get the relations \( \lom^3 = R_E \los^2 \) and 
\( R_E = \tilde{g} \los \) using the standard relations
$f^{-1/2} R_E = g_s l_s$ and $l_p^3 = f^{-1/2} R_E l_s^2$ 
with $R_E$ being the radius of the electric circle.
The relation \( \lom^3 = R_E \los^2 \) means that the open string
in NCOS theory comes from the open membrane in OM theory
wrapped on the electric circle.

\section{${\rm NCOS}_{5+1}$ theory and $(1,1)$ OBLST/OD1-theory}

\subsection{Why new theories?}

Some of the most important reasons that we need to consider the
possibility of new theories, are

\begin{itemize}

\item As we shall see shortly, there are evidence for dynamical 
closed strings in the theory defined by the ${\rm NCOS}_{5+1}$
decoupling limit. These closed strings corresponds to the 
'little' strings of ordinary LST, and they have a different tension than
the open strings. This means we need to have a new theory 
that can have both an open string phase and a closed string phase.

\item As has been considered 
in \cite{Gopakumar:2000ep,Bergshoeff:2000ai,Kawano:2000gn}, 
the ${\rm NCOS}_{p+1}$
theories for $p \leq 4$ can be obtained as the dimensional
reduction of OM theory,
just as ${\rm SYM}_{p+1}$ theories for $p \leq 4$ can
be obtained as the dimensional reduction of $(2,0)$ SCFT.
But the ${\rm SYM}_{5+1}$ and $(2,0)$ SCFT are connected through 
the $(2,0)$ and $(1,1)$ LSTs which reduce at low energies to
$(2,0)$ SCFT and ${\rm SYM}_{5+1}$, respectively, and which are
connected by T-duality.
The question is if there are a similar type of connection between
${\rm NCOS}_{5+1}$ and OM theory through two new Little String Theories
having also open branes in their spectra.

\item If we consider thermal ${\rm NCOS}_{5+1}$ theory 
we can also ask what the high energy behavior of this can be.
As was seen in \cite{Harmark:2000wv,Klebanov:2000pp,Gubser:2000mf} 
the temperature can exceed the Hagedorn temperature of the NCOS theory.
Therefore, the high energy behavior of ${\rm NCOS}_{5+1}$ theory 
cannot be described by weakly coupled NCOS theory.

\end{itemize}

All this points to the existence of two new theories generalizing 
the $(1,1)$ and $(2,0)$ LST, with 'little' strings.
These new theories should be T-dual and should reduce to ${\rm NCOS}_{5+1}$
and OM theory when the 'little' strings are heavy.

\subsection{S-duality on the ${\rm NCOS}_{5+1}$ limit}

Consider the ${\rm NCOS}_{5+1}$ limit of the F1-D5 bound state
as in Section \ref{NCOSsec}
\begin{equation}
\label{F1D5limit1}
\bls \rightarrow 0
\spa
\eps = \frac{\bls^2}{\los^2}
\spa
\bgs = \tilde{g} \eps^{-1}
\end{equation}
\begin{equation}
\label{F1D5limit2}
g_{\mu \nu} = \eps^{-1} \eta_{\mu \nu} 
\spa
\mu,\nu = 0,1
\spa
g_{ij} = \eps \delta_{ij} 
\spa
i,j = 2,...,9
\end{equation}
\begin{equation}
B_{01} = \frac{1}{\eps} \frac{1}{2\pi \bls^2} - \frac{1}{4\pi \los^2}
\spa
A_{2345} = \frac{1}{(2\pi)^3 \tilde{g} \los^4}
\end{equation}
Doing an S-duality transformation, this corresponds to
the D1-NS5 bound state in the limit
\begin{equation}
\label{D1NS5limit1}
g_s \rightarrow 0
\spa
l_s = \mbox{fixed}
\spa
\eps = \frac{g_s l_s^2}{\los^2} = g_s \tilde{g}
\end{equation}
\begin{equation}
\label{D1NS5limit2}
g_{\mu \nu} = \eps^{-1} \eta_{\mu \nu} 
\spa
\mu,\nu = 0,1
\spa
g_{ij} = \eps \delta_{ij} 
\spa
i,j = 2,...,9
\end{equation}
\begin{equation}
A_{01} = \frac{1}{\eps} \frac{1}{2\pi \bls^2} - \frac{1}{4\pi \los^2}
\spa
A_{2345} = \frac{1}{(2\pi)^3 \tilde{g} \los^4}
\end{equation}
where \( l_s^2 = \bgs \bls^2 \) and \( g_s = 1 / \bgs \).
This limit of the D1-NS5 bound state gives an 
equivalent description of the theory since the limits are S-dual.

\subsection{'Little' closed strings}

As have already been described, the ${\rm NCOS}_{5+1}$ theory
has open strings of tension $\frac{1}{4\pi \los^2}$ 
\cite{Seiberg:2000ms,Gopakumar:2000na}.
We are now ready to list the evidence for closed strings
in the theory defined by the two S-dual decoupling limits 
\eqref{F1D5limit1}-\eqref{F1D5limit2} 
and \eqref{D1NS5limit1}-\eqref{D1NS5limit2}:

\begin{itemize}

\item ${\rm NCOS}_{5+1}$ reduces to ${\rm SYM}_{5+1}$ 
for low energies, and ${\rm SYM}_{5+1}$ has closed string solitons
of tension $\frac{(2\pi)^2}{\gym^2} = \frac{1}{2\pi l_s^2}$.

\item In the limit $\los \rightarrow 0$ and $l_s = \mbox{fixed}$ 
the theory on D1-NS5 reduces to the usual LST on the NS5-brane
with dynamical 'little' strings.

\item As we review in Section \ref{SGthermo}, the thermodynamics calculated
via the supergravity dual predicts Hagedorn behavior with
a Hagedorn temperature given by the 'little' closed string scale.
This means that the 'little' closed strings are dynamical.

\end{itemize}

We conclude that the theory on the F1-D5/D1-NS5 bound state
contains 'little' closed strings of tension $\frac{1}{2\pi l_s^2}$.
We therefore have a new theory which encompass both ${\rm NCOS}_{5+1}$
and $(1,1)$ LST \cite{Harmark:2000ff,Gopakumar:2000ep}.
This theory is called $(1,1)$ open brane Little String Theory (OBLST)
in \cite{Harmark:2000ff} since it is almost the same as
$(1,1)$ LST, the only difference is that the d0-particle
in $(1,1)$ LST is replaced by the open string of ${\rm NCOS}_{5+1}$.
In \cite{Gopakumar:2000ep} it is called open D1-brane (OD1) theory
since in the D1-NS5 decoupling limit, the open brane origins from
a D-string. More precisely, the effective tension of 
an open D-string stretching along the electrical field is
\( T_{\rm eff} = \frac{1}{\eps} \frac{1}{2\pi g_s l_s^2} - A_{01}
= \frac{1}{4\pi \los^2} \).
The connection between the two string lengths and the 
NCOS coupling is \( l_s^2 = \tilde{g} \los^2 \).

\subsection{Supergravity dual, thermodynamics and phase diagrams}
\label{SGthermo}

The supergravity dual of $(1,1)$ OBLST/OD1-theory is found using the limit
\eqref{F1D5limit1} on the F1-D5 bound state and the limit 
\eqref{D1NS5limit1} on the D1-NS5, were the rescaling of the metric 
is replaced by an equivalent scaling of the coordinates.
This is done in \cite{Harmark:2000ff}.
As shown in \cite{Harmark:2000wv,Harmark:2000ff,Harmark:2000qm} 
the thermodynamics computed from the F1-D5/D1-NS5 supergravity
dual of $(1,1)$ OBLST/OD1-theory is 
\begin{equation}
T = \tlst 
\spa
F = 0
\spa 
S = \frac{1}{\tlst} E
\spa
E = \frac{V_5}{(2\pi)^5} \frac{\tilde{r}_0^2}{\los^4 l_s^4}
\end{equation}
with
\begin{equation}
\tlst = \frac{1}{2\pi l_s \sqrt{N}}
\end{equation}
where $T$ is the temperature, $F$ is the free energy, $E$ is the energy
and $S$ is the entropy. 
Moreover, $V_5$ is the five-volume and $\tilde{r}_0 = \frac{r_0}{\sqrt{\eps}}$
where $r_0$ is the horizon radius of the non-extremal supergravity solution.
This thermodynamics exhibits Hagedorn behavior with Hagedorn 
temperature $\tlst$ as noticed in 
\cite{Harmark:2000hw,Harmark:2000ff,Harmark:2000qm}
for this case and in 
\cite{Maldacena:1996ya,Maldacena:1997cg,Harmark:2000hw,Berkooz:2000mz} 
for the ordinary LSTs.

Clearly, when the supergravity dual is a valid description of 
$(1,1)$ OBLST/OD1-theory
we have \( T \sim \tlst \).
Moreover, the 'little' closed strings dominates when the supergravity
dual is valid since the Hagedorn temperature $\tlst$ is given
by the 'little' string scale \cite{Harmark:2000ff}.

For the ordinary LSTs it was found in \cite{Harmark:2000hw,Berkooz:2000mz}
that the entropy as function of temperature has the 
critical behavior
\begin{equation}
S(T) \propto (\tlst - T)^{-1}
\end{equation}
for \( (\tlst - T) / \tlst \ll 1/N^3 \). This was found by adding 
one-loop string corrections to the temperature.
In \cite{Harmark:2000hw,Harmark:2000wv,Harmark:2000ff,Harmark:2000qm}
the critical behavior for OD1-theory was instead found to be
\begin{equation}
S(T) \propto (\tlst - T)^{-2/3}
\end{equation}
for \( (\tlst - T) / \tlst \ll 1/N^3 \). This was found by doing
tree-level string corrections to the temperature.

We see that the $(1,1)$ LST and the $(1,1)$ OBLST/OD1-theory
have different critical behavior.
This means that even when the 'little' strings are the dominant
degrees of freedom the two theories are different.
This we interpret to origin from having a different geometry
which in the $(1,1)$ OBLST/OD1-theory should be a 
space-time non-commutative geometry with \( [x^0,x^1] = i 2\pi \los^2 \).

From the above we see that we have four phases of $(1,1)$ OBLST/OD1-theory:
\begin{itemize}
\item ${\rm SYM}_{5+1}$
\item ${\rm NCOS}_{5+1}$
\item $(1,1)$ LST
\item NC $(1,1)$ LST
\end{itemize}
where 'NC $(1,1)$ LST' means the 
phase of $(1,1)$ OBLST/OD1-theory where the 'little' strings 
dominates but the geometry supposedly is space-time non-commutative.

We now consider the phase diagrams of $(1,1)$ OBLST/OD1-theory.
In the phase diagrams, we have defined the energy variable $u$
as $u = \frac{1}{\los^2} \frac{r}{\sqrt{\eps}} $
where $r$ is the radial coordinate of the supergravity solution.

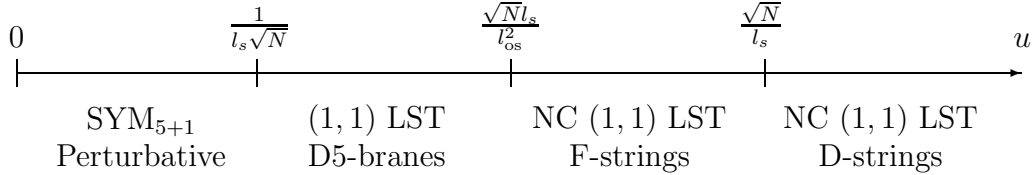
\begin{figure}[ht]
\begin{picture}(390,65)(0,0)
\put(35,40){\vector(1,0){380}}
\put(412,50){$u$}
\put(50,5){\shortstack{${\rm SYM}_{5+1}$ \\ Perturbative}}
\put(145,5){\shortstack{$(1,1)$ LST \\ D5-branes}}
\put(230,5){\shortstack{NC $(1,1)$ LST \\ F-strings}}
\put(325,5){\shortstack{NC $(1,1)$ LST \\ D-strings}}
\put(32,50){0}
\put(35,35){\line(0,1){10}}
\put(115,55){$\frac{1}{l_s \sqrt{N}}$}
\put(126,35){\line(0,1){10}}
\put(210,55){$\frac{\sqrt{N} l_s}{\los^2}$}
\put(222,35){\line(0,1){10}}
\put(308,55){$\frac{\sqrt{N}}{l_s}$}
\put(318,35){\line(0,1){10}}
\end{picture}
\caption{Phase diagram for $(1,1)$ OBLST/OD1-theory with 
\( 1/N \ll \tilde{g} \ll 1 \). \label{figOB11b} }
\end{figure}

The first case is $1/N \ll \tilde{g} \ll 1$, depicted 
in figure \ref{figOB11b}. 
In this case the NCOS theory is strongly coupled \( \tilde{g} N \gg 1 \)
and the 'little' strings are light in the sense that
\( \frac{1}{N} \frac{1}{2\pi l_s^2} \ll \frac{1}{2\pi \los^2} \).
The scale $\frac{1}{N} \frac{1}{2\pi l_s^2}$ is not directly the
'little' string scale but it is the scale connected with the
Hagedorn temperature $\tlst$ and is therefore a scale that enters
in the dynamics of 'little' strings (see \cite{Harmark:2000hw,Harmark:2000ff}
for further discussion). 
We see that there are not any NCOS phase, which is sensible since it is 
strongly coupled and the NCOS scale is relatively heavy.

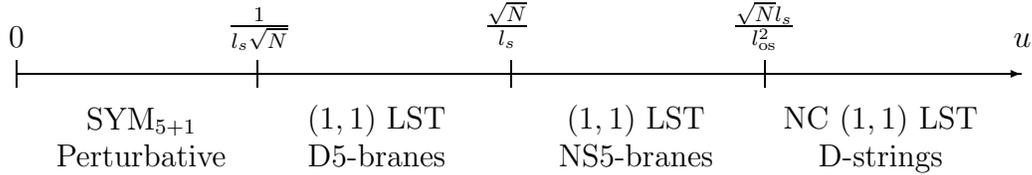
\begin{figure}[ht]
\begin{picture}(390,65)(0,0)
\put(35,40){\vector(1,0){380}}
\put(412,50){$u$}
\put(50,5){\shortstack{${\rm SYM}_{5+1}$ \\ Perturbative}}
\put(145,5){\shortstack{$(1,1)$ LST \\ D5-branes}}
\put(240,5){\shortstack{$(1,1)$ LST \\ NS5-branes}}
\put(325,5){\shortstack{NC $(1,1)$ LST \\ D-strings}}
\put(32,50){0}
\put(35,35){\line(0,1){10}}
\put(115,55){$\frac{1}{l_s \sqrt{N}}$}
\put(126,35){\line(0,1){10}}
\put(212,55){$\frac{\sqrt{N}}{l_s}$}
\put(222,35){\line(0,1){10}}
\put(306,55){$\frac{\sqrt{N}l_s}{\los^2}$}
\put(318,35){\line(0,1){10}}
\end{picture}
\caption{Phase diagram for $(1,1)$ OBLST/OD1-theory with 
\( \tilde{g} \gg 1 \). \label{figOB11c} }
\end{figure}

The second case is \( \tilde{g} \gg 1 \), depicted in figure \ref{figOB11c}.
Here we clearly have strong NCOS coupling \( \tilde{g} \gg 1 \)
and the NCOS scale is large compared to the 'little' string scale
\( \frac{1}{2\pi l_s^2} \ll \frac{1}{2\pi \los^2} \).
Thus, we do not have any NCOS phase but instead a SYM phase, a LST
phase and a non-commutative LST phase, which again is sensible since
the 'little' strings are much lighter than the NCOS theory strings,
and the NCOS theory is strongly coupled.
 
\begin{figure}[ht]
\begin{picture}(390,65)(0,0)
\put(35,40){\vector(1,0){380}}
\put(412,50){$u$}
\put(65,5){\shortstack{${\rm NCOS}_{5+1}$ \\ Perturbative}}
\put(180,5){\shortstack{NC $(1,1)$ LST \\ F-strings}}
\put(305,5){\shortstack{NC $(1,1)$ LST \\ D-strings}}
\put(32,50){0}
\put(35,35){\line(0,1){10}}
\put(150,55){$\frac{1}{\los}$}
\put(157,35){\line(0,1){10}}
\put(274,55){$\frac{\sqrt{N}}{l_s}$}
\put(284,35){\line(0,1){10}}
\end{picture}
\caption{Phase diagram for $(1,1)$ OBLST/OD1-theory with 
\( \tilde{g} \ll 1/N \). \label{figOB11a} }
\end{figure}
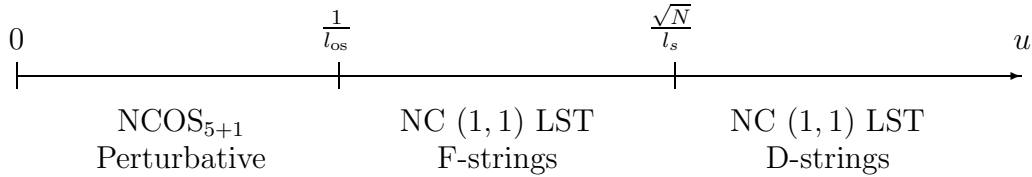

The final case is \( \tilde{g} \ll 1/N \), depicted in figure \ref{figOB11a}.
Here we have a NCOS phase for low energies, which is sensible
since the NCOS coupling is weak \( \tilde{g} N \ll 1 \) and since
the NCOS scale is light 
\( \frac{1}{2\pi \los^2} \ll \frac{1}{N} \frac{1}{2\pi l_s^2} \).
But, we still enter a non-commutative LST phase at high energies.
This can be explained by the fact that 
the Hagedorn temperature of NCOS theory is $\tncos \sim \frac{1}{\los}$
so that $\tncos \ll \tlst$. This means that 
there must be an NCOS theory Hagedorn transition before $u$ reaches 
$\frac{1}{\los}$ since for $u \gg \frac{1}{\los}$ we have $T \sim \tlst$.
Thus, the non-commutative LST phase comes after the NCOS theory has been
subject to a Hagedorn transition.

\section{OM theory and $(2,0)$ OBLST/OD2-theory}

\subsection{Motivation and definition}

Consider the $(1,1)$ OBLST/OD1-theory. This is obtained from the
D1-NS5 bound state. But T-dualizing this we get D2-NS5, which
is nothing but M2-M5 with a compact transverse direction.
Thus, defining a T-dual theory of the $(1,1)$ OBLST/OD1-theory
in this way we can get a theory that can reduce to OM-theory 
for low energies. This provides a nice connection via dualities
between ${\rm NCOS}_{5+1}$ and OM theory.

Thus, we define a new theory by the OM theory limit 
\eqref{OMlimit1}-\eqref{OMlimit2} on the M2-M5 bound-state with a transverse
circle of radius $R_T$. 
When $R_T$ is small we can use a type IIA description
in terms of a D2-NS5 bound state in the limit
\begin{equation}
\label{OD2limit1}
\hat{g}_s \rightarrow 0
\spa
l_s = \mbox{fixed}
\spa
f = \frac{\hat{g}_s^2 l_s^6}{\lom^6}
\end{equation}
\begin{equation}
\label{OD2limit2}
g_{\mu \nu} = f^{-1} \eta_{\mu \nu} 
\spa
\mu,\nu = 0,1,2
\spa
g_{ij} = f \delta_{ij} 
\spa
i,j = 3,...,9
\end{equation}
\begin{equation}
A_{012} = \frac{1}{f} \frac{1}{(2\pi)^2 l_s^3} - \frac{1}{2(2\pi)^2 \lom^3}
\spa
A_{345} = \frac{1}{(2\pi)^2 \lom^3}
\end{equation}
This new theory on the D2-NS5 bound state is called $(2,0)$ OBLST
in \cite{Harmark:2000ff}
since it is almost like $(2,0)$ LST, but with an open membrane 
instead of the d1-brane in $(2,0)$ LST.
In \cite{Gopakumar:2000ep} it is called open D2-brane (OD2) theory
since the open membrane origins from an open D2-brane.
The open membrane has the tension
\( T_{\rm eff} = \frac{1}{f} \frac{1}{(2\pi)^2 g_s l_s^3} - A_{012}
= \frac{1}{2(2\pi)^2 \lom^3} \).
The $(2,0)$ OBLST/OD2-theory has 'little' strings of tension
$\frac{1}{2\pi l_s^2}$.

\subsection{T-duality to $(1,1)$ OBLST/OD1-theory}

We can now consider doing a T-duality transformation
on $(2,0)$ OBLST/OD2-theory in an electric direction.
Choose the direction of $x^2$ to be on a circle of radius
$R_E$. 
Using the standard T-duality transformation
\begin{equation}
g_s = \hat{g}_s \frac{l_s}{f^{-1/2} R_E}
\spa
R_E R_E' = l_s^2
\spa
g_{22} g_{22}' = 1
\end{equation}
we see from the limits \eqref{D1NS5limit1}-\eqref{D1NS5limit2} and 
\eqref{OD2limit1}-\eqref{OD2limit2} that the 
T-duality works if and only if \( f = \eps \) and 
\( \lom^3 = R_E \los^2 \).
From this we get
\begin{equation}
2\pi R_E \frac{1}{2 (2\pi)^2 \lom^3} = \frac{1}{4\pi \los^2}
\end{equation}
so that the tension of the open string is equal to the tension of the
wrapped open membrane.
Thus, the open membrane and the open string are T-dual to each other.
This provides a new connection between ${\rm NCOS}_{5+1}$ and 
OM theory in the sense that their fundamental objects are T-dual.

\subsection{Phase diagrams}

The thermodynamics of $(2,0)$ OBLST/OD2-theory computed from 
the supergravity dual plus string correction terms is the same as
the one reviewed for $(1,1)$ OBLST/OD1-theory in \ref{SGthermo}.

We have again four different phases
\begin{itemize}
\item $(2,0)$ SCFT 
\item OM theory
\item $(2,0)$ LST
\item NA $(2,0)$ LST
\end{itemize}
'NA $(2,0)$ LST' refers to the phase of 
$(2,0)$ OBLST/OD2-theory where the 'little' strings dominate
but the geometry supposedly is non-associative.
We get two phase diagrams, depicted in figure
\ref{figOB20a} and \ref{figOB20b}, where $\tilde{r} = r / \sqrt{f}$
with $r$ being the radial coordinate in the M2-M5/D2-NS5 supergravity
solution \cite{Harmark:2000ff} (see also \cite{Alishahiha:2000er}).

\begin{figure}[ht]
\begin{picture}(390,65)(0,0)
\put(35,40){\vector(1,0){380}}
\put(412,50){$\tilde{r}$}
\put(50,5){\shortstack{$(2,0)$ SCFT \\ ${\rm AdS}_7 \times S^4$}}
\put(145,5){\shortstack{OM theory \\ M2-branes}}
\put(235,5){\shortstack{$(2,0)$ OBLST \\ M2-branes}}
\put(330,5){\shortstack{$(2,0)$ OBLST \\ D2-branes}}
\put(32,50){0}
\put(35,35){\line(0,1){10}}
\put(110,55){$N^{1/3} \lom $}
\put(126,35){\line(0,1){10}}
\put(216,55){$R_T$}
\put(222,35){\line(0,1){10}}
\put(305,55){$\frac{\sqrt{N} \lom^6}{\l_s^5}$}
\put(318,35){\line(0,1){10}}
\end{picture}
\caption{Phase diagram for $(2,0)$ OBLST/OD2-theory. \label{figOB20a} }
\end{figure}

\begin{figure}[ht]
\begin{picture}(390,65)(0,0)
\put(35,40){\vector(1,0){380}}
\put(412,50){$\tilde{r}$}
\put(50,5){\shortstack{$(2,0)$ SCFT \\ ${\rm AdS}_7 \times S^4$}}
\put(145,5){\shortstack{$(2,0)$ LST \\ M5-branes}}
\put(239,5){\shortstack{$(2,0)$ LST \\ NS5-branes}}
\put(330,5){\shortstack{NA $(2,0)$ LST \\ D2-branes}}
\put(32,50){0}
\put(35,35){\line(0,1){10}}
\put(120,55){$R_T$}
\put(126,35){\line(0,1){10}}
\put(209,55){$\frac{\sqrt{N} \lom^3}{l_s^2}$}
\put(222,35){\line(0,1){10}}
\put(305,55){$\l_s \sqrt{N}$}
\put(318,35){\line(0,1){10}}
\end{picture}
\caption{Phase diagram for $(2,0)$ OBLST/OD2-theory. \label{figOB20b} }
\end{figure}

\section{Other OD$p$-theories}

The OD1 and OD2-theories were discovered 
in \cite{Harmark:2000ff,Gopakumar:2000ep}, but in \cite{Gopakumar:2000ep} 
there were in addition discovered other OD$p$-theories. 
The OD$p$-theories, $p=0,1,...,5$, are defined via the limit%
\cite{Gopakumar:2000ep}
\begin{equation}
\label{ODp1}
l_s \rightarrow 0
\spa
\eps = \frac{l_s^4}{\leff^4}
\spa
g_s = \tilde{g} \eps^{\frac{3-p}{4}}
\end{equation}
\begin{equation}
\label{ODp2}
g_{\mu \nu} = \eta_{\mu \nu} 
\spa
\mu,\nu = 0,1,...,p
\spa
g_{ij} = \eps \delta_{ij} 
\spa
i,j = p+1,...,9
\end{equation}
\begin{equation}
A_{01\cdots p} = \frac{1}{(2\pi)^p g_s l_s^{p+1}}
- \frac{1}{2(2\pi)^p \tilde{g} \leff^{p+1}}
\spa
A_{p+1 \cdots 5} = \frac{1}{(2\pi)^{4-p} \tilde{g} \leff^{5-p}}
\end{equation}

The OD$p$-theories have 'little' strings of tension 
$\frac{1}{2\pi \leff^2}$ and open $p$-branes of tension
\( T_{\rm eff} = \frac{1}{(2\pi)^p g_s l_s^{p+1}} - A_{01\cdots p}
= \frac{1}{2(2\pi)^p \tilde{g} \leff^{p+1}} \).
The supergravity duals of OD$p$-theories are given in 
\cite{Alishahiha:2000pu,Harmark:2000qm} for all $p=0,1,...,5$.
Their thermodynamics computed from supergravity plus string 
corrections have been studied in
\cite{Harmark:2000qm} for all $p=0,1,...,5$.

\vskip0.5cm
\noindent
{\large \bf Acknowledgments}

\smallskip
\noindent
We thank the organizers for a nice and fruitful conference.
We thank Niels Obers for collaboration and many useful discussions.
We also thank Jan de Boer and Robbert Dijkgraaf for 
many interesting discussions and comments.


\end{document}